\RequirePackage{fix-cm}
\documentclass[smallextended]{svjour3}       % onecolumn (second format)
\smartqed  % flush right qed marks, e.g. at end of proof

\usepackage{url}
\usepackage[a4paper]{geometry}
\usepackage{amssymb}
\usepackage{euscript}
\usepackage{graphicx}
\graphicspath{{images/}}
\usepackage{color}
\usepackage{epsfig}
\usepackage{ulem}
\newcommand{\image}[3]{
\begin{figure}[#1]
\begin{center}
\includegraphics{full_#2.pdf}
\caption{\small#3}
\label{image:#2}
\end{center}
\end{figure}
}

\newcommand*{\doi}[1]{\href{https://doi.org/\detokenize{#1}}{DOI: \detokenize{#1}}}

\def\BRF{B_{\mbox{\scriptsize RF}}}
\def\TNS{T_{\mbox{\scriptsize NS}}}
\def\THE{T_{\mbox{\scriptsize He}}}
\def\sh{\delta_{\omega}}

\usepackage[colorlinks=true,hypertexnames=true]{hyperref}
\begin{document}

%\title{New oscillation modes of HPD in superfluid $^3$He-B}
\title{Spectroscopy of oscillation modes in homogeneously precessing domain of superfluid $^3$He-B}

%\author{V.~V.~Zavjalov\/\thanks{e-mail: v.zavjalov@lancaster.ac.uk},
% A.~Savin, E.~Sergeicheva, P.~J.~Hakonen}
\author{V.~V.~Zavjalov\/\thanks{e-mail: v.zavjalov@lancaster.ac.uk; pertti.hakonen@aalto.fi} \and
 A.~Savin \and E.~Sergeicheva \and P.~J.~Hakonen}
 

\titlerunning{Spectroscopy of HPD oscillation modes}        % if too long for running head


\institute{V.~V.~Zavjalov \and
 A.~Savin \and E.~Sergeicheva \and P.~J.~Hakonen 
\at Low Temperature Laboratory, Aalto University School of Science, Finland
 \at Department of Applied Physics, Aalto University School of Science, Finland
% \at Department of Physics, Lancaster University, UK
 %              \\
 %          \email{pertti.hakonen@aalto.fi}           %  
           \\
\emph{Present address:} of V.~V.~Zavjalov   %  if needed
\at Department of Physics, Lancaster University, UK
}

\date{\today}
\maketitle

\begin{abstract}
We study Homogeneously Precessing Domain (HPD) in superfluid $^3$He-B in
a regular continuous-wave nuclear magnetic resonance (CW NMR) experiment.
Using Fourier analysis of CW NMR time traces, we identify several oscillation modes with frequency monotonically increasing with the frequency shift of the HPD. Some of these modes are localized near
the cell walls, while others are localized in bulk liquid and can be interpreted as
oscillations of $\vartheta$-solitons. We also observe chaotic motion of the HPD
in a certain range of temperatures and frequency shifts.
\end{abstract}

%\subsection*{Introduction}
\section{Introduction}

Nuclear magnetic resonance (NMR) in superﬂuid $^3$He has played an important role in studies of its properties since the discovery of superfluid phases in 1972~\cite{1972_he3_super}. The condensate in superfluid $^3$He is formed by p-wave pairing of fermions (pairs having spin 1 and orbital angular momentum 1), which allows the order parameter of the system to be written as 3x3 complex matrix. A few superﬂuid phases with distinct broken symmetries have been observed~\cite{VolovikBook}; the most studied phases are the A phase and the B phase~\cite{VollhardtWolfle}. In the B phase, the order parameter structure involves an arbitrary 3D rotation matrix which sets the mutual orientation of spin and orbital spaces. The degrees of freedom of this matrix lead to three possible spin-wave modes, strongly affected by spin-orbit interaction. As a result, many unique phenomena can be observed in NMR experiments, including signatures of non-uniform spatial distribution and topological defects in the order-parameter ﬁeld~\cite{Hakonen1989,1976_maki_solitons,Maki1977}, longitudinal NMR~\cite{longitudinalNMR}, and Homogeneously Precessing Domain (HPD)~\cite{1985_hpd_e}.

The HPD was discovered in 1985~\cite{1985_hpd_e,1985_hpd_f_e}. Since then, it has been used as a convenient probe in experiments with spin supercurrents~\cite{1987_phase1} and vortices~\cite{1991_rotb_nonaxisym}. Various oscillation modes of HPD have been studied, including uniform rotational oscillations around magnetic field~\cite{2005_he3b_hpd_osc,2012_kupka_hpd}, corresponding non-uniform oscillations~\cite{2008_skyba_osc}, and oscillations of the HPD surface~\cite{1992_hpd_osc,1997_lokner_hpd,2003_gazo_hpd}.

One of possible 2D topological defects in $^3$He-B is the so-called $\vartheta$-soliton. It appears between two regions where the order-parameter rotation matrix stays in different configurations at the minimum of spin-orbit interaction energy. In the regime of linear NMR, the analysis of the $\vartheta$-soliton is quite straightforward~\cite{1976_maki_solitons}. The solitons can also be present in the HPD state~\cite{1992_mis_hpd_topol}, but their structure is more involved. $\vartheta$-solitons in HPD have been observed experimentally in conjunction with spin-mass vortices in Ref.~\cite{1992_prl_kondo_smv}. Recently, dynamics of a $\vartheta$-soliton in the HPD has been considered in numerical simulations \cite{2022_zav_num}. In this work, we experimentally observe spatially localized HPD oscillation modes that can be compared with simulations and attributed to oscillations of the $\vartheta$-soliton, which so far have remained unidentified in experiments.

%\subsection*{Theoretical background}
\section{Theoretical background}
In the B phase of superfluid~$^3$He, broken symmetries of the
order parameter can be described by a 3x3 matrix
\begin{equation}
A_{aj} = R_{aj}({\bf\hat n}, \vartheta)e^{i\varphi},
\end{equation}
where $R_{aj}$ is a rotation matrix with axis~$\bf\hat n$ and
angle~$\vartheta$, and $e^{i\varphi}$ is a complex phase factor. Various spatial distributions of $\varphi$ and~$R_{a j}$ are possible, as well as a few
types of topological defects. Spin dynamics of $^3$He-B is
described by non-linear Leggett equations~\cite{1975_he3_teor_leggett,VollhardtWolfle}
for spin~$\bf S$ and matrix~$R_{aj}$:
\begin{eqnarray}\label{eq:legg}
\dot S_a &=&
  [{\bf S}\times \gamma {\bf B}]_a 
  + \frac{4}{15}\,\frac{\Omega_B^2\chi}{\gamma^2}\ \sin\vartheta(4\cos\vartheta + 1)\ n_a
  - \nabla_k J_{ak},
\\\nonumber
\dot R_{aj} &=&
  e_{abc} R_{cj} \Big(\frac{\gamma^2}{\chi} {\bf S} - \gamma {\bf B} \Big)_b.
\end{eqnarray}
Here $\gamma$ is the gyromagnetic ratio of~$^3$He, $\bf B$ denotes the external
magnetic field, $\chi$ is the magnetic susceptibility, $e_{abc}$ is
a permutation tensor, $J_{ak}$ specifies the spin current which carries component $a$
of spin in the direction $k$, and $\Omega_B$ is the Leggett frequency, a measure
of spin-orbit interaction strength in superfluid $^3$He. All relaxation terms are neglected for
simplicity.

Homogeneously precessing domain is a solution of these non-linear equations~\cite{1985_hpd_f_e}.
Consider a coordinate system having axis~$\bf\hat z$ along the magnetic field,
and rotating around this direction with some frequency $\omega \ge \gamma B$. Then, the
HPD state is given by
\begin{equation}\label{eq:hpd}
{\bf n} = {\bf\hat y}
,\qquad
\cos\vartheta = -\frac{1}{4} - \frac{15}{16}\,\frac{\omega(\omega-\gamma B)}{\Omega_B^2}
,\qquad
{\bf S} = \frac{\omega\chi}{\gamma^2}
  \left[{\bf\hat x}\sin\vartheta  + {\bf\hat z}\left(\cos\vartheta - \frac{\omega-\gamma B}{\omega}\right) \right].
\end{equation}
Note that this solution can have an arbitrary
orientation around ${\bf\hat z}$ axis because the system is symmetric,
but in a regular continuous-wave nuclear magnetic resonance (CW NMR)
experiment, the presence of a radio-frequency field along ${\bf\hat x}$ axis
stabilizes HPD in ${\bf n}\parallel{\bf\hat y}$ orientation in the rotating frame.
The important parameter here is frequency shift $\sh=\omega-\gamma B$.
Usually, the shift is small, i.e. angle $\vartheta$ is slightly bigger than the so-called Leggett angle,
$\vartheta_L = \cos^{-1}(-1/4)\approx104^\circ$, and the spin is tilted by approximately
the same angle in the direction of the axis $\bf\hat x$. The fact that the deflection
of the spin is connected with the precession frequency $\omega$ makes an HPD state stable
in a non-uniform magnetic field. If some spatial gradient appears in the precession
frequency, spin currents transfer magnetization and compensate the
difference. As a result, homogeneous precession takes place in the whole
volume where the HPD exists. In fact, the HPD behaves like a liquid which
fills low-field parts of the experimental cell up to the level $\omega=\gamma
B$. In the case of free precession, this level is determined by the total
system energy which decreases because of relaxation. In a continuous-wave NMR experiment with sufficient feed of energy, the level can be controlled by the pumping frequency or magnetic field.

%\ph{Soliton as a domain wall between two minima - how this works in the case of HPD? This paragraph should be made clearer.} 
Topology of the HPD state is more complicated than that of an equilibrium state
of~$^3$He-B because, in addition to the orbital order parameter distribution, one
can have a non-trivial distribution of spin. One possible 2D topological
defect in this state is the $\vartheta$-soliton~\cite{1992_mis_hpd_topol}. Minimum of spin-orbit interaction energy in $^3$He-B is achieved at the Leggett angle $\vartheta_l = \cos^{-1}(-1/4)$, the soliton appears between two different energy minima, $\vartheta_L$ and $2\pi - \vartheta_L$. There are two characteristic length scales in the $\vartheta$-soliton in HPD. First, a small core region of size $\xi_D\approx
10$~$\mu$m, across which the angle $\vartheta$ changes; $\xi_D$ is referred to as ``the dipolar length'' because its scale is governed by the dipolar energy in comparison to gradient terms. Second, outside the core region, the magnetization varies on the
length scale~$\xi_\omega$ that is inversely proportional to square root of the
frequency shift~\cite{1992_mis_hpd_topol}:
\begin{equation} \label{xi}
\xi_\omega  = \frac{c_\parallel}{\sqrt{\omega(\omega - \gamma B)}}
\end{equation}
where $c_\parallel$ is spin-wave velocity. Typical range of frequncy shifts in our experiments is $10-100$~Hz. This corresponds to $\xi_\omega=0.15-0.5$~mm.

An analytical calculation of $\vartheta$-solitons in HPD is quite an involved task. In Ref.~\cite{2022_zav_num}, we have performed numerical simulations in one-dimensional geometry, to obtain the structure and dynamics of the $\vartheta$-soliton in HPD. A few low-frequency modes were identified in the simulation, and their dependencies on temperature, frequency shift, and radio-frequency field were determined. We use the simulated signatures and their characteristics to identify the measured oscillation modes in this work.

%\subsection*{Experiment}
\section{Experiment}
Our experiments were performed on a pulse-tube-based nuclear
demagnetization cryostat with minimum temperature of 0.2 mK~\cite{2014_drydemag}.
%The experiment presented here was done in Aalto University, in a dry
%demagnetization cryostat~\cite{2014_drydemag} in 2018. 
The experimental chamber
was made using epoxy Stycast-1266 (see Fig.~1). It has cylindrical shape with
an inner diameter of $7.8$~mm and a length of $9$~mm. The experimental volume is connected to the heat exchanger of the nuclear stage through a channel having a diameter of 1~mm in its most narrow section.

%%%%%%%%%%%%%%%%%%%%%%
\image{ht}{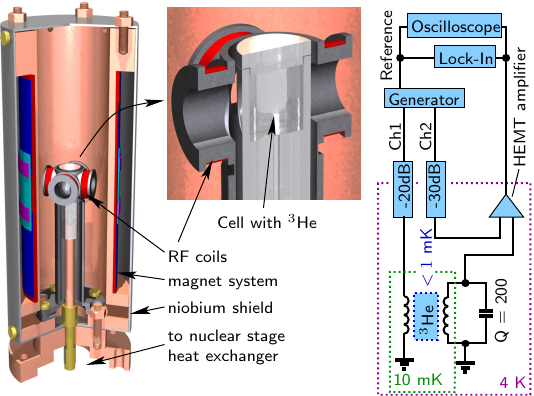}{Experimental cell with NMR magnet (left), close up of the NMR coils and the cylindrical volume (center), and schematics of the NMR spectrometer (right). Detailed description of the setup is given in the text.}
%%%%%%%%%%%%%%%%%%%%%%

The magnet system for uniform static field includes a solenoid with a bore diameter of 36~mm, a gradient coil and a quadratic field
coil. The magnet system, thermally anchored to the mixing chamber of the cryostat, is surrounded by superconductive niobium shield. By contact to the nuclear stage, this shield
provides mechanical stabilization, while keeping thermal isolation between the magnet
system and the demagnetization cooling stage. The field homogeneity in the cell volume was
measured using NMR linewidth in normal $^3$He. By adjusting current in the gradient and
quadratic field coils to their optimal values, it was possible to achieve
a homogeneity of~$1.5\cdot10^{-4}$ over the sample volume. These normal-state NMR measurements also gave us a calibration
of the gradient coil: $0.157$ (T/m)/A.

% main: calculated: 398 G/A, measured: 405.9 G/A
% grad: calculated: 31.4 G/cm/A, measured: 15.7 G/cm/A,  grad0 = -7.5mA
% RF: Vexc=1V, Vcomp=5.727V (see http://slazav.xyz:8085/get_prev?name=nmr_gen&t2=1528810318)
% on the coil:
%  Vexc = 1V / 2(pk2pk->amp) / 10 (20db) -> 0.05V
%  Vcomp = 5.727V / 2(pk2pk->amp) / sqrt(1000) (30db) -> 0.09055V
% each coil inductance L = 55.5 uH, frequency 1.124Hz -> Z = 391.96 R
% Hexc  = 0.05V / 391.96R  * 1.66 mT/A  -> 211.76 nT  -> 2.11 mG ->  6.87 Hz
% Hcomp = 0.09055V / 391.96R * 1.66 mT/A -> 383.49 nT -> 3.83 mG -> 12.44 Hz
% Hrf = (Hexc + Hcomp)/2 = 297.62 nT -> 2.976 mG -> 9.653 Hz

Our NMR spectrometer consists of two pairs of RF-coils, aligned perpendicular to each other, located symmetrically at a distance of 7~mm from the central axis of the cell. These RF-coils with a diameter of 12~mm are made of 50~$\mu$m copper wire.
The inductance of each coil pair is 55.5~$\mu$H, and the calculated value of the RF-field
in the center is $1.66$~mT/A. One coil pair is used for applying NMR
excitation from a signal generator, while the other one is embedded into an $LC$ tank
circuit with a resonant frequency of $1.124$~MHz and $Q=200$. The signal from
the receiving coils is amplified by a home-made HEMT
amplifier~\cite{2019_zav_amp}. Differential input of the amplifier is used
to compensate for the background signal received away from the NMR resonance. By using the excitation and
compensation voltages and the parameters of the electric circuit, one
can calculate currents in both coil pairs and the total amplitude of RF
field. At an RF-excitation of 1~V$_{pp}$ (on the generator output), we need compensation voltage of 5.73~V$_{pp}$ which corresponds to the RF field rms amplitude $\BRF=297.6$~nT. The NMR signal is recorded by a lock-in amplifier tuned at the excitation frequency and, in parallel, by a digital oscilloscope.

Our experiments were done at a pressure of 25.7~bar in a magnetic field
of 34.67~mT (the NMR frequency equals to the resonant frequency of the tank
circuit). Cooling capacity of the cryostat allowed us to measure a few hours
in superfluid~$^3$He in each demagnetization cycle.
Measurements were done both during demagnetization to the lowest temperature
and while warming up.

Temperature was measured using a SQUID-based noise thermometer attached to the
nuclear stage. Since it is possible to observe the superfluid transition temperature $T_c$ by means
of NMR, we measured temperature difference between liquid helium and the nuclear stage at temperature $T_c$ in the sample volume. The result can be described using a simple model with
a thermalization time~$\tau_0$: $(\THE - K\,\TNS)/\tau_0 = -  K\,\dot\TNS$,
where $\THE$ and $\TNS$ are temperatures of the helium volume and the nuclear stage, respectively,
and $K\simeq 1$ is an adjustment factor for the noise thermometer calibration. Normally the noise thermometer is calibrated at higher temperature against thermometer at the mixing chamber, the factor $K$ fixes inaccuracy of this calibration. Measurements at different rates of cooling give us a time constant $\tau_0=1540$~s. A rough estimation of the thermalization time using Kapitza resistance $R_K = 900/T$~[K$^2$m$^2$/W]~\cite{1984_franco_sinter} and normal $^3$He heat
capacity $C_N=34.5T$~[J/K/mol]~\cite{1983_greywall_norm_he3} for our amount of helium
(approximately 1 mole) and sinter area~($20$~m$^2$) give a quite similar value $\tau_0=1550$~s. 

Below the superfluid transition temperature, the Kapitza resistance is expected to follow $1/T$ law~\cite{1978_ahonen_kap_res}, on the basis of which the thermalization time can be estimated as~$\tau = C_B/C_N\,\tau_0$ where $C_N$ and $C_B$ denote the heat capacities of
normal phase and B phase of $^3$He, respectively. By integrating the heat transfer
model, we may estimate the temperature of~$^3$He, $\THE$, during the whole experiment as a function of time. The use of $\THE$ significantly reduces the observed
hysteresis in temperature-dependent frequency shifts measured during cool-down
and warm-up, which indicates that our simple model works well.

There is also another thermometric uncertainty: during HPD measurements with strong NMR excitation we have a significant (about 0.2~$T_c$) overheating of $^3$He sample.
It arises because of the narrow channel between the HPD volume and the heat exchanger volume. A correction of this overheating is implemented in a simple fashion (see Fig.~2A): we determine temperature-dependent frequencies of HPD oscillations (details are given later) and assume that they should extrapolate to zero at $T_c$. We do extrapolation using a smooth rational function and obtain a temperature shift, which is naively assumed to be temperature independent. We do this separately for each sequence of measurements, because different measurement parameters (RF-field amplitude, gradient, range of magnetic field sweep) result in different overheating. The correction has been applied to all given values of $T$ except those in Fig. 2A where uncorrected value $\THE$ is used. Accuracy of such temperature corrections and our thermometry in general is quite poor, of the order of 0.1$T_c$. Nevertheless this does not affect main results of the work. 

The HPD in our experiments was created in the usual continuous-wave method by sweeping magnetic field in the presence of large enough RF-pumping at the resonance frequency of the $LC$ circuit. The sweeping rate was always
0.807~$\mu$T/s (26.2~Hz/s in frequency units). The appearance of the HPD could easily be
 observed in the NMR signal recorded by the lock-in amplifier. In parallel,
we recorded the same signal by oscilloscope to observe the response in a wide range
of frequencies. Oscillations of HPD can be seen as modulation of the main
signal, with side bands separated from the main signal by the frequency of the
oscillations. The oscillations were visible without any special
excitation. We did measurements as a function of temperature, field
gradient $\nabla B$, and RF-field amplitude $\BRF$. Because of
overheating we were able to create HPD only in a small range of RF-fields and only at small field gradients. 
Note that field homogeneity of the NMR magnet $1.5\cdot10^{-4}$ corresponds to field variations of
about $5\,\mu$T across the cell, while measurements were done at the absolute value of applied gradient from 0 to $20\,\mu$T/cm. This means that the actual field profile and the process of HPD growth in our experiment are not well-defined. There could be a few field minima and maxima in the cell which lead to a complicated topology of the HPD surface, and this could be a possible source of $\vartheta$-solitons. However, the main results in this work are obtained on a well-defined coherent HPD that is obtained in the regime where the HPD fills the whole cell. 

We were unable to observe uniform oscillations of HPD~\cite{2005_he3b_hpd_osc,2012_kupka_hpd}, presumably because we could not apply a rapid step change in $B$ to excite them. It would have been useful to have a separate longitudinal coil around the cell to excite uniform oscillations and use their frequency for better temperature and $\BRF$ calibration. 

An important experimental parameter is the frequency shift $\sh = \omega - \gamma B$, the difference between pumping frequency $\omega$ which is always constant, and the Larmor frequency proportional to magnetic field $\bf B$, which changes during the experiments.
The frequency shift varies across the sample because of the field inhomogeneity and the applied field gradient $\nabla{\bf B}$. The inhomogeneity limits the accuracy of calibration of $B$ and the absolute value of the frequency shift $\sh$. However, relative changes in the frequency shift are well-defined and accurate.
Normally, we calibrate the magnetic field using NMR in liquid helium above $T_c$, and calculate the frequency shift using this calibration. We can only say that, at zero frequency shift condition, $\omega = \gamma B$ is valid in some region inside the sample chamber. Such frequency shift values are presented in Figs.~2, 3, and 4. For high-temperature modes (displayed in Fig.~5), we use a more specific calibration of the magnetic field: we assume
that the field at which the modes appear corresponds to zero frequency shift at the exact position of the soliton (details will be given in Results section).

%%%%%%%%%%%%%%%%%%%%%%

\image{ht}{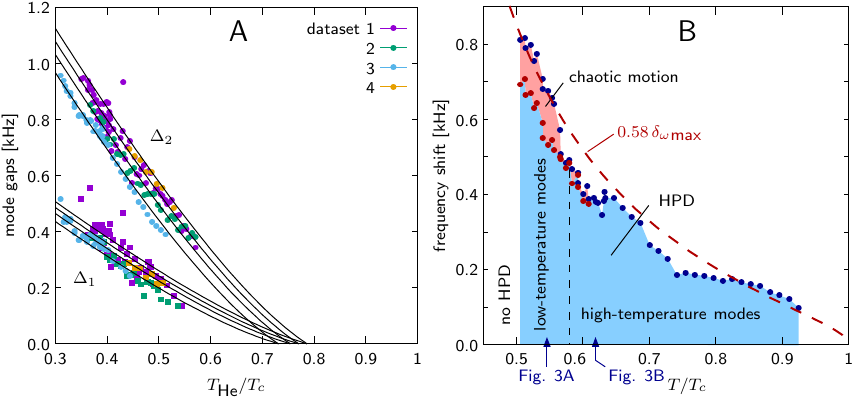}{
{\bf A:} Overheating correction for HPD measurements. We measure
high-frequency mode gaps (will be explain later in Results section) as a
function of $\THE$ temperature, assume that they are proportional to each
other, and extrapolate them to zero in order to find the actual $T_c$ in the sample volume. Four different data sets are
displayed, taken at different excitation levels. 
{\bf B:} Typical range of corrected temperatures and frequency shifts over which we
detect HPD using $\BRF=357$~nT (1.2~V$_{pp}$) and a field
gradient of $\nabla B=-0.94\,\mu$T/cm. The dashed red line is a theoretical curve
for the maximum frequency shift using Eq.~(\ref{eq:dmax}), with the value scaled down by a factor of 0.58 to match the measured data. }

% (-8.1mA) - (-7.5mA) * 1.57 mT/cm/A
%%%%%%%%%%%%%%%%%%%%%%

%\subsection*{Results}
\section{Results}
Fig.~2B displays typical range of temperatures and frequency shifts, over which we
can observe the HPD state. During each measurement, the magnetic field is
swept continuously upwards, the HPD appears when frequency shift crosses zero in some part of the cell and exists until a maximum frequency shift is reached. We found that this maximum value
can be estimated in the following way. Using Leggett's equations with
radio-frequency field $\BRF$ and including the Leggett-Takagi relaxation term~\cite{1977_leggett_takagi}, one can find the steady state that corresponds to the HPD. In the presence of Leggett-Takagi relaxation, the vector $\bf
n$ is deflected from the direction given by Eq.~(\ref{eq:hpd}), in such a way that 
\begin{equation}
n_x = \frac{\sqrt{15}}{4}\,
      \frac{\sh^2}{\Omega_B^2}\,
      \frac{1}{\gamma \BRF}\,
      \frac{1}{\tau}
\end{equation}
where $\tau$ is the Leggett-Takagi relaxation time which is of the order of the
mean Bogolyubov quasiparticle relaxation time (we use values calculated in Ref.~\cite{1978_einzel_transp_rel}). We do not know exact conditions at which the HPD becomes unstable, but we can say that it should happen before the vector $\bf n$ is rotated by 90 degrees, i.e. $n_x\approx 1$. Due to these approximations we can not have an exact expression, but only an 
estimation for the maximum frequency shift:
\begin{equation}\label{eq:dmax}
\sh{}_{\mbox{max}}
  \approx \Omega_B \sqrt{\gamma \BRF\,\tau}
\end{equation}
In Fig.~2B, we display this estimation which matches the experimental results quite well when scaled by a factor of $0.58$. Furthermore, the good agreement in temperature dependence indicates that our thermometry corrections described in the Experiment section are realistic.

%%%%%%%%%%%%%%%%%%%%%%
\image{ht}{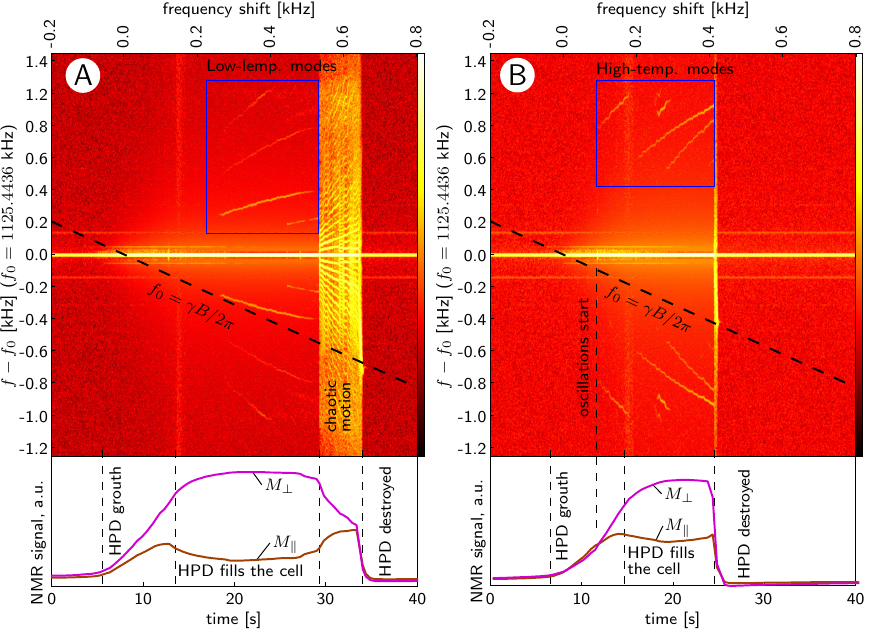}{Typical recorded signals.
Upper pictures display spectrograms
obtained by Fourier transform on time-domain magnetization signals captured by a digital oscilloscope. The vertical axis is the frequency (difference
from constant NMR frequency~$f_0$), horizontal axis is time (bottom axis)
or frequency shift (upper axis), which is changing because of constant
ramping of magnetic field, below them are signals measured with lock-in
amplifier at the excitation frequency.
{\bf A:} $T=0.54\,T_c$, HPD with chaotic region and ``low-temperature modes''
(side bands marked with a blue frame).
{\bf B:} $T=0.62\,T_c$, ``high-temperature modes'' (marked with a blue frame).
Data in A and B were taken at pressure 25.7 bar, $\BRF=357$~nT, and $\nabla B = -0.94$~$\mu$T/cm.}
%%%%%%%%%%%%%%%%%%%%%%

We find an HPD state at temperatures above approximately $0.5\,T_c$. Below this
temperature the HPD is unstable because of catastrophic relaxation effects
(\cite{1989_europhys_ipp_catrel,2007_jltp_surovtsev_suhl}). We distinguish two
temperature ranges, below and above $T \approx 0.59\,T_c$, with completely
different HPD oscillation modes. We call them as ``low-temperature''
and ``high-temperature'' modes. At low temperatures and high frequency
shifts we also observe chaotic motion of the HPD. This effect was studied
numerically by Y. Bunkov in Ref.~\cite{1993_bunk_chaot}. All these features can
be seen in the spectrograms displayed in Fig.~3. Frequency spectrum is plotted as a function of time (bottom scale) and frequency shift corresponding to the change of magnetic field in frequency units (upper scale). The left plot of Fig.~3A illustrates data recorded
at $0.54\,T_c$ on the low-temperature modes and chaotic motion of the HPD;
the right plot Fig.~3B presents data of the high-temperature modes recorded at
$0.62\,T_c$. All modes are seen as symmetric side bands of the NMR frequency $\omega$.
From the measured signals, we extract mode frequencies $\Omega$ as a function of the
frequency shift $\sh$. We can see a clear dependence of mode frequencies on the frequency shift
in the presence of field inhomogeneity and different values of the applied field gradient. This
indicates that the oscillation modes are localized.

{\bf Low-temperature modes} are seen at $T < 0.59\,T_c$. We observe two different modes and their harmonics at $2\times$, $3\times$ the fundamental frequency (see the blue frame in Fig. 3A).
Mode frequencies are proportional to the square root of the frequency shift and can be written as
\begin{equation}\label{eq:lmodes}
\Omega^2 = S_{LT}\, (\sh-\sh^0),
\end{equation}
where the slope $S$ and offset $\sh^0$ are mode-dependent parameters as seen for the data in Fig.~4A. We can assume that the low-temperature modes do
not have any intrinsic frequency gap and the offset~$\sh^0$ is determined
only by the location inside the cell in the presence of a field gradient. We found
that~$\sh^0$ does not depend on temperature nor on the amplitude of the RF-field, but it does
depend on the field gradient in a way as if the modes are localized at the upper
and lower ends of the experimental cell (Fig.~4B). The temperature and
RF-field dependence of the slope parameter~$S_{LT}$ of mode frequencies is illustrated in Fig.~4C.

%%%%%%%%%%%%%%%%%%%%%%
\image{ht}{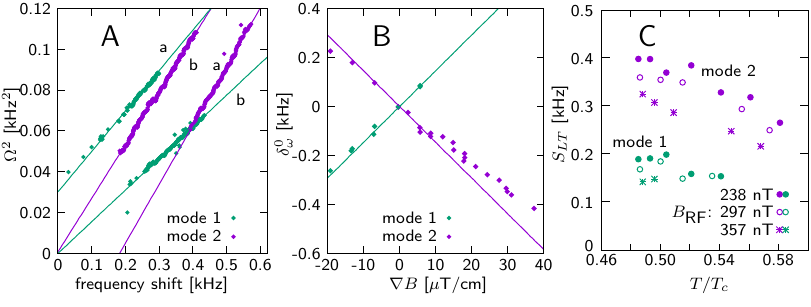}{Characteristics of low-temperature modes.
{\bf A:} Example of mode frequencies as a function of the frequency shift,
extracted from two signals measured using two different field
gradients: (a) $\nabla B = -13$~$\mu$T/cm, (b) $\nabla B = -0.6$~$\mu$T/cm. Both signals
contain two modes which are indicated in green (mode 1) and magenta (mode 2). One can see linear dependence of $\Omega^2$ on $\sh$. Both signals have been measured with
$\BRF=297$~nT and temperature $T=0.58\,T_c$.
{\bf B:} Mode offset position~$\sh^0$ as a function of the field gradient.
Solid lines are frequency shifts at cell ends calculated using the known
field profile of the gradient coil and the cell size. Signals were
measured with $\BRF=298$ and 358~nT in the whole temperature range
over which the low-temperature modes were visible.
{\bf C:} Temperature dependence of mode slopes $S_{LT}$. Different symbols are
used for signals measured with different RF-field amplitudes:
($\bullet$), ($\circ$), ($\ast$) correspond to $\BRF=238,297,357$~nT,
respectively. $\nabla B = -0.6$~$\mu$T/cm. The two colors are the same as in the other panels for the two different modes.
}
%%%%%%%%%%%%%%%%%%%%%%

{\bf High-temperature modes.} At higher temperature we observe a different
set of oscillation modes (see the blue frame in Fig.~3B). In some rare cases
they coexist with low-temperature modes, but usually there is a clear
transition between these two regimes. There are many high-temperature
modes, some are more common, while some appear only in a few measurements.
Examples of two measurements are shown in Fig.~5, different modes are marked by numbers 1\dots8.

\image{h}{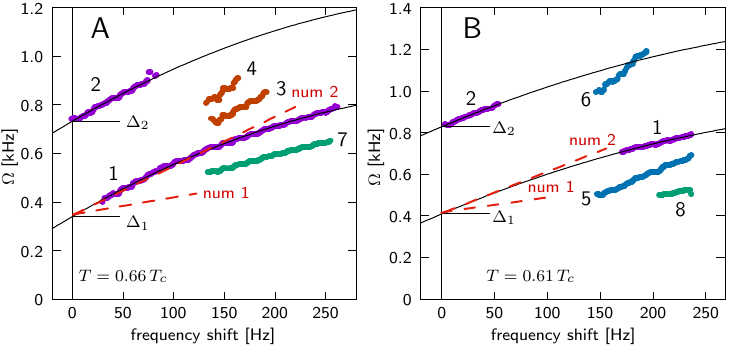}{
Examples of high-temperature modes. $\BRF=297$~nT, $\nabla B=-0.9\,\mu$T/cm.
The frequency shift is counted from the point where mode 2 appears, as explained in the text. {\bf A: $T=0.66\,T_c$},
{\bf B: $T=0.61\,T_c$}.
Different modes are marked by numbers 1\dots8.
Black lines is a quadratic fit of modes 1 and 2 with a single frequency-shift dependence and different gaps.  Red dashed lines ``num 1'' and ``num 2'' are results of numerical simulations in Ref.~\cite{2022_zav_num}.
}

Modes 1 and 2 are the most common ones, they appear in almost all measurements in the high-temperature regime. If we try to extrapolate their frequencies to
zero, like we did with low-temperature modes, we find a big negative frequency shift
which can not be explained by field inhomogeneity. It means that the modes have some intrinsic frequency gap. We found that at the point where mode 2 becomes visible, a tiny but
clear feature on the HPD signal can be observed ("oscillations start" line
in Fig.~3B). We assume that this corresponds to the localization point of
the modes, and the frequency shift should be measured from this value. We find that this point is always located in a random place
inside the cell (we can convert the magnetic field value to a position using the field gradient). This allows us to determine the mode gaps~$\Delta_1$ and~$\Delta_2$. We see
that the modes 1 and 2 have exactly the same frequency shift dependence,
but there is a constant separation specified by the difference of $\Delta_1$ and $\Delta_2$. (see Fig. 5).

This information leads us to formulate our central conjecture: We reason
that the high-temperature modes are oscillations of a $\vartheta$-soliton localized
somewhere in the cell volume. Low-frequency oscillations of the $\vartheta$-soliton are calculated 
using 1D geometry as discussed in Ref. ~\cite{2022_zav_num}. In our
experiment, the soliton is comprised of a 3D circular membrane, in which case the radial part of oscillations should give an additional frequency gap
\begin{equation}\label{eq:vel}
\Delta_n = a_n C/r,
\end{equation}
where $C$ is the wave velocity along the membrane, $r$ is the radius, and $a_n$ denote
the zeros of the Bessel function of the first kind. For two first radial modes
$a_1=2.405$, $a_2=5.520$. Note that Ref.~\cite{2022_zav_num} discusses another type of
oscillation of the circular membrane, namely in which the whole
soliton is moving. These modes should have much larger frequency, but we do
not find them in the experiment. In Fig.~5, the dashed lines marked as ``num 1'' and ``num
2'' display results of the 1D numerical simulation, given by the approximative formulas Eqs. (9) and (10)
of Ref.~\cite{2022_zav_num}, shifted by $\Delta_1$. Apart from the gaps $\Delta_1$ and $\Delta_2$, we see that the measured modes 1 and 2 behave closely to the second calculated mode for the $\vartheta$-soliton. The behavior observed for modes 7 or 8 could correspond to the first mode, but it's hard to conclude with certainty owing to the small visibility range of the mode in our experiment.

Further basic information on the high-temperature modes 1 and 2 is presented in Fig.~6. The mode gaps~$\Delta_1$ and~$\Delta_2$ are plotted as a function of temperature in Fig.~6A. These data are the same as in Fig.~2A, but with the temperature
correction applied. We do not observe any noticeable dependence of the gaps on
$\BRF$ and $\nabla B$. In Fig.~6B, the ratio $\Delta_2/\Delta_1$ is plotted
for measurements where both modes exist. The observed ratio is pretty close to the theoretical ratio given by
$a_2/a_1 = 2.295$. The values deduced for the wave velocity $C = \Delta_i r / a_n$ are
depicted in Fig.~6C. On Fig.~6D we compare linear term in the mode frequency $S_{HT} = \frac{d\Omega}{d\sh}(\sh=0)$ with numerical simulation. We find good agreement with the calculated second mode of soliton oscillations, with only a weak dependence on temperature and $\BRF$. The fact that we can a separate temperature-dependent gap and an almost temperature-independent soliton mode supports our choice of $\sh=0$ for the frequency shift scale.

\image{ht}{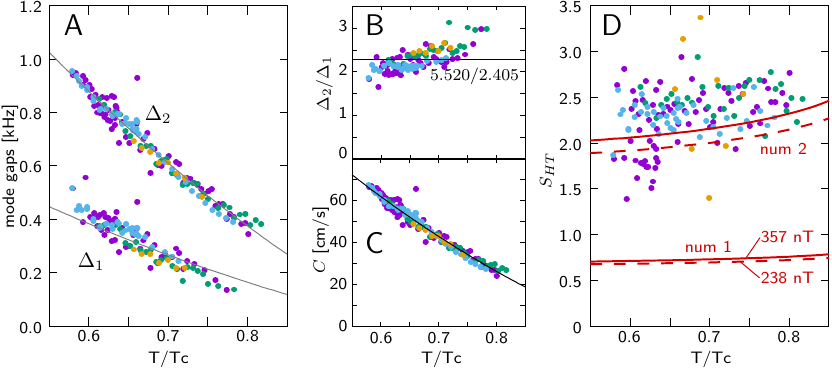}{
High-temperature modes 1 and 2 vs. $T$. Colors refer to different data sets
measured in different cooling cycles using $\BRF=238,297,357$~nT and $\nabla B =-0.9\,\mu$T/cm.
{\bf A:} Gaps~$\Delta_1$ and~$\Delta_2$ vs. temperature. The data is the same
as in Fig.~2A but now with temperature correction. 
{\bf B:} Ratio $\Delta_2/\Delta_1$ for measurements where both modes exist.
{\bf C:} Wave velocity $C$ evaluated from $\Delta_1$ and $\Delta_2$ using Eq.~(\ref{eq:vel}).
{\bf D:} Slope $S_{HT}$ of mode frequency $\Omega$ vs. frequency shift $\sh$ in comparison with numerical simulations (lines ``num1'' and ``num2''). Dashed lines are calculated using $\BRF=238$~nT and the solid ones using
$\BRF=357$~nT.
}

We leave detailed analysis of the other high-temperature modes outside the scope of this paper.
Modes 3 and 4 (see Fig.~5A) were quite common, but they existed only in a small
frequency shift range. They look like the next two modes of $\vartheta$-soliton
oscillations obtained in the numerical simulation, but at the moment we
can not confirm it. Modes 5 and 6 (see Fig.~5B) appeared in pairs with a frequency ratio of 2. They do not extrapolate to the same energy gap as modes 1 and
2, and probably they do not have a gap. 
Modes 7 and 8 were visible only in a few
signals, which is unfortunate as they seem to bear resemblance with the first mode in our simulations. However, due to scarce information we could not analyze and classify
them accurately.

%%%%%%%%%%%%%%%%%%%%%%

%\subsection*{Conclusions}
\section{Conclusions}
In this work, we measured the response of the HPD state in about 10~kHz band
around the NMR frequency and could make interesting observations: new
oscillation modes localized near the cell walls, additional modes in the bulk volume of the HPD that we attribute to oscillations of $\vartheta$-solitons, and
chaotic motion at high frequency shifts. We also estimated the maximum
frequency shift at which the HPD state is stable and found that our experimental result matches our estimations based on Leggett-Takagi relaxation.

The richness of the observed soliton oscillation modes makes the most interesting finding of our experiments. 
Our work shows that, overall, the observed mode frequencies match with numerical simulations, but there are still more
questions than answers: how these modes are excited, how the solitons are
created, how to classify them and explain all the observed modes. Further
investigation is certainly needed, in experiment, theory, and numerical
simulations. It would be interesting to study how solitons are created at
different magnetic field ramping rates, to isolate a single soliton by
tuning experimental conditions without destroying the HPD, to populate certain
modes by using additional RF-field excitation pulses, etc.
It would be important to study how creation of a soliton is affected by
magnetic field profile and sweeping rate.
Altogether, the soliton dynamics would be a
sophisticated and quite fundamental problem for further investigations.

%\subsection*{Acknowledgements}
\section*{Acknowledgements}
 This work was supported by the Academy of Finland projects 341913 (EFT) and 312295, 352926 (CoE, Quantum Technology Finland) as well as by ERC (grant no. 670743). The research leading to these results has received funding from the European Union’s Horizon 2020 Research and Innovation Programme, under Grant Agreement no 824109. The experimental work benefited from the Aalto University OtaNano/LTL infrastructure.  We would like to thank Vladimir Dmitriev, Yury Bunkov, and
Grigory Volovik for useful discussions.

\section*{Data availability}
Our measured data are available on 10.5281/zenodo.8431264. It includes oscilloscope records, spectrograms, and extracted oscillation modes for about six hundred measurements. 
%It is possible that some new information can be extracted from these data.
\bibliographystyle{spphys}
\bibliography{paper.bib}

%\printbibliography
\end{document}